\documentclass[aps, superscriptaddress, nofootinbib, preprintnumbers]{revtex4}

\usepackage{amsmath}	% AMS Math Package
\usepackage{amsthm}		% Theorem Formatting
\usepackage{amssymb}	% Math symbols such as \mathbb
\usepackage{eufrak}
\usepackage{datetime}
\usepackage{graphicx}
\usepackage{color}
\usepackage{verbatim}
\usepackage{bm}
\usepackage[colorlinks=true, citecolor=midblue, linkcolor=midblue, urlcolor=midblue]{hyperref}

\usepackage{setspace}

\usepackage{fancyhdr}

\definecolor{grey}{rgb}{0.4,0.4,0.4}
\definecolor{dullmagenta}{rgb}{0.4,0,0.4}
\definecolor{darkblue}{rgb}{0,0,0.4}
\definecolor{midblue}{rgb}{0,0,0.7}
\definecolor{midred}{rgb}{0.5,0,0}
\definecolor{orange}{rgb}{1,0.5,0}
\definecolor{lightbrown}{rgb}{0.75,0.5,0.25}
\definecolor{tan}{cmyk}{0.14,0.42,0.56,0}
\definecolor{djunglegreen}{cmyk}{0.99,0,0.52,0}
\definecolor{lightgreen}{rgb}{0,1,0}
\definecolor{olivegreen}{cmyk}{0.64,0,0.95,0.40}
\definecolor{midgreen}{rgb}{0.0,0.675,0.0}
\definecolor{darkgreen}{rgb}{0,0.5,0}

\newcommand{\vs}{\vspace}

\renewcommand{\.}{\hspace{0.5mm}}

\newcommand{\ra}{\ensuremath{\rightarrow}}

\newcommand{\ral}{\ensuremath{\longrightarrow}}

\newcommand{\Hrm}{\ensuremath{\mathrm{H}}}

\newcommand{\Rrm}{\ensuremath{\mathrm{R}}}

\newcommand{\Vrm}{\ensuremath{\mathrm{V}}}

\newcommand{\crm}{\ensuremath{\mathrm{c}}}

\newcommand{\erm}{\ensuremath{\mathrm{e}}}
\newcommand{\frm}{\ensuremath{\mathrm{f}}}

\newcommand{\srm}{\ensuremath{\mathrm{s}}}

\newcommand{\Ocal}{\ensuremath{\mathcal{O}}}
\newcommand{\Pcal}{\ensuremath{\mathcal{P}}}

\newcommand{\Scal}{\ensuremath{\mathcal{S}}}

\newcommand{\Cbb}{\ensuremath{\mathbb{C}}}

\renewcommand{\d}{\ensuremath{\mathrm{d}}}
\smallskipamount
\settimeformat{ampmtime}

\usepackage{float}

\fancypagestyle{plain}{ %
  \fancyhf{} % remove everything
}

\begin{document}

%%%%%%%%%%%%%%%%%%%%%%%%%%%%%%%%%%%%%%%%%%%%%%%%%%%%%%%%%
\title{Primordial Black Holes with Multimodal Mass Spectra}

\author{Bernard Carr}
\email{b.j.carr@qmul.ac.uk}
\affiliation{Department of Physics and Astronomy,
	Queen Mary University of London,
	Mile End Road,
	London E1 4NS,
	United Kingdom}
	
\author{Florian K{\"u}hnel}
\email{kuhnel@kth.se}
\affiliation{Department of Physics,
	School of Engineering Sciences,\\
	KTH Royal Institute of Technology,
	AlbaNova University Center,\\
	Roslagstullsbacken 21,
	106\.91 Stockholm,
	Sweden}

\affiliation{The Oskar Klein Centre for Cosmoparticle Physics,
	AlbaNova University Center,\\
	Roslagstullsbacken 21,
	106\.91 Stockholm,
	Sweden}

\date{\formatdate{\day}{\month}{\year}, \currenttime}

%%%%%%%%%%%%%%%%%%%%%%%%%%%%%%%%%%%%%%%%%%%%%%%%%%%%%%%%%
\begin{abstract}
A mechanism for generating primordial black-hole mass spectra with many spikes is proposed and investigated. This mechanism relies on the choice of non-Bunch-Davies vacua, these leading to oscillatory features in the inflationary power spectrum. This in turn generates oscillations in the primordial black-hole mass function with exponentially enhanced spikes. This ``multimodal'' effect is demonstrated for most of the well-studied models of primordial black-hole formation.
\end{abstract}

%%%%%%%%%%%%%%%%%%%%%%%%%%%%%%%%%%%%%%%%%%%%%%%%%%%%%%%%%
\maketitle

%%%%%%%%%%%%%%%%%%%%%%%%%%%%%%%%%%%%%%%%%%%%%%%%%%%%%%%%%
\section{Introduction}
\label{sec:Introduction}

After $50$ years, there is still no definite evidence for primordial black holes (PBHs). However, there are numerous ways in which they could have formed in principle, so even their non-existence places important constraints on models of the early Universe. For this reason attention has often focussed on {\it limits} on the number of PBHs in various mass ranges \cite{Carr:2009jm}. These limits extend from the Planck mass of $10^{-5}\.$g to around $10^{12}\.M_{\odot} \sim 10^{45}\.$g and therefore span some $50$ decades of mass. A distinction is usually made between PBHs smaller than $M_{*} \sim 10^{15}\.$g, which have already evaporated through Hawking radiation \cite{Hawking:1974rv}, and those larger than $M_{*}$, which would still exist. In the wake of Hawking's discovery, attention initially focussed on evaporating PBHs but in recent years it has shifted to non-evaporating ones. Indeed, PBHs larger than $M_{*}$ are often invoked to {\it explain} certain observational features{\;---\;}such as the presence of dark matter \cite{Carr:2016drx}, the black-hole coalescences detected by LIGO \cite{Bird:2016dcv}, the supermassive black holes in galactic nuclei \cite{Bean:2002kx} and certain features of large-scale cosmological structure \cite{Carr:2018rid}.

In most PBH formation scenarios one expects the PBHs to span a range of masses \cite{Carr:2016drx}. If they form from large-amplitude primordial inhomogeneities, the shape of the mass function, $\d n / \d M$, depends on the form of the inhomogeneities. If they are scale-invariant (ie.~with constant amplitude when they fall inside the particle horizon), then the spectrum has a simple power-law form, $\d n / \d M \propto M^{- \alpha}$, with an exponent which depends on the equation of state \cite{Carr:1975qj}. This also applies if the PBHs form form the collapse of cosmic strings \cite{Hawking:1987bn}. In the simplest inflationary scenario, the fluctuations will be nearly (but not exactly) scale-invariant and one expects $\d n / \d M$ to have a lognormal form \cite{Dolgov:1992pu}. In more complicated inflationary scenarios, the generated fluctuations may exhibit a feature on some scale, in which case the PBHs may only form over a narrow mass range. If the PBHs form through critical collapse, most of them will have the horizon mass but there will also be a low-mass tail of PBHs with $\d n / \d M$ having a steep positive power-law form \cite{Yokoyama:1998qw}. During an early matter-dominated era, PBHs form much more easily \cite{Khlopov:1980mg, 1982SvA....26..391P} and one expects $\d n / \d M \propto M^{-2}$ over a wide range of masses for nearly scale-invariant primordial fluctuations. 

The fact that the PBHs are likely to have an extended mass function has important implications for the interpretation of the PBH constraints. Typically, one derives constraints on $f( M )$, the fraction of the dark matter in PBHs with mass around $M$, on the assumption that they have a monochromatic mass function (ie.~a mass range $\Delta M$ comparable to $M$). Various approaches have been used for the extended case but the analysis is quite complicated and one cannot directly compare the predicted form of $f( M )$ with the monochromatic limits \cite{Carr:2016drx}. This is because the limits are themselves modified for an extended mass function \cite{Carr:2017jsz}. If one wants the PBHs to play some {\it combination} of cosmological r{\^o}les (eg.~providing the dark matter and the LIGO coalescences), then the form of the mass function is particularly crucial.

An extended PBH mass function often has a single peak, in which case it can be described by two parameters (the position and width of the peak). For a power-law mass function, the dominant contribution to the density comes from the upper or lower cut-off rather than some intermediate mass but two parameters still suffice. However, in principle, the PBH mass function could be more complicated than this. Indeed, with a contrived choice of the inflation potential $\Vrm( \phi )$, the power spectrum of the primordial fluctuations could have an arbitrary form \cite{Ivanov:1994pa}{\;---\;}with several peaks in the power spectrum{\;---\;}and this would generate corresponding peaks in the PBH mass spectrum. We will describe this as a {\it multimodal} mass spectrum. For example, it has been shown that this may arise due to oscillations in the sound speed \cite{Cai:2018tuh}. 

In this paper, we discuss another context which arises fairly naturally during inflation and necessarily generates a multimodal mass spectrum. This arises because of a modulation of the primordial power spectrum generated by stringy or trans-Planckian effects \cite{Bergstrom:2002yd, ArmendarizPicon:2003gd}. The power spectrum has a number of spikes, corresponding to the harmonics of some fundamental length-scale. On a large scale this could be observed in the cosmic microwave background (CMB) or through periodicity in large-scale structure observations \cite{Einasto:1997md}. However, it could also affect PBH production on small scales. This is because the probability of PBH formation is exponentially sensitive to the ratio of the density perturbation to the critical value required for PBH formation. For a blue (red) primordial spectrum, only the highest (lowest)-$k$ peak is important. For a flat spectrum, all the peaks could be important but whether one can see the higher-$k$ ones depends upon the relative width of the spikes and the spacing between them.

The plan of this paper is as follows. In Sec.~\ref{sec:PBH-Formation} we briefly review various possible PBH formation scenarios. In Sec.~\ref{sec:Vacua-2} we examine how the choice of non-Bunch-Davies vacua affects the primordial power spectrum. A discussion of resonance effects follows in Sec.~\ref{sec:Resonance-Effects}, their influence on the modulation of the power spectrum being similar. In Sec.~\ref{sec:PBH-Scenarios} we apply the results of Sec.~\ref{sec:Vacua-2} to the PBH scenarios presented in Sec.~\ref{sec:PBH-Formation}. We draw some conclusions in Sec.~\ref{sec:Conclusion-and-Outlook}.

%%%%%%%%%%%%%%%%%%%%%%%%%%%%%%%%%%%%%%%%%%%%%%%%%%%%%%%%%
\section{Unmodulated PBH Mass Function}
\label{sec:PBH-Formation}

\noindent In many scenarios, one would expect PBHs to form with an extended mass function. In this section, we discuss five such scenarios. These assume that the PBHs form from scale-invariant primordial fluctuations or the collapse of cosmic strings (Sec.~\ref{sec:Collapse-from-Scale--Invariant-Fluctuations-or-Cosmic-Strings-II}), an early matter-dominated era (Sec.~\ref{sec:Collapse-in-a-Matter--Dominated-Era-II}), initial inhomogeneities of inflationary origin (Sec.~\ref{sec:Collapse-from-Inflationary-Fluctuations-II}), critical collapse (Sec.~\ref{sec:Critical-Collapse-II}) or the softening of the equation of state at the QCD phase transition (Sec.~\ref{sec:QCD-Phase-Transition-II}). In each case, we will give the form of the mass function and the dark-matter fraction. In order to understand the implications of the effect discussed in this paper, we need to understand how bumps in the power spectrum affect the PBH mass function $\d n / \d M$ and the current dark-matter fraction $f( M )$. Whether the power spectrum spikes generate distinct mass-function spikes depends upon the ratio of the width of the mass-function spikes to their separation.

%%%%%%%%%%%%%%%%%%%%%%%%%%%%%%%%%%%%%%%%%%%%%%%%%%%%%%%%%
\subsection{Collapse from Scale-Invariant Fluctuations or Cosmic Strings}
\label{sec:Collapse-from-Scale--Invariant-Fluctuations-or-Cosmic-Strings-II}

If the PBHs form from scale-invariant fluctuations (ie.~with constant amplitude at the horizon epoch), their mass spectrum should have the power-law form \cite{Carr:1975qj}
\begin{align}
	\frac{ \d n }{ \d M }
		&\propto
				M^{-\alpha}
	\quad
	\mathrm{with}
	\quad 
	\alpha
		=
				\frac{2\.(1 + 2 \gamma ) }{ 1 + \gamma }
				\; ,
				\label{eq:spectrum}
\end{align}
where $\gamma$ specifies the equation of state ($p = \gamma\.\rho\.c^{2}$) at PBH formation. The exponent arises because the background density and PBH density have different redshift dependencies. The mass function is also proportional to the probability $\beta( M )$ that an overdense region of mass $M$ has a size exceeding the Jeans length at maximum expansion, so that it can collapse against the pressure. In this case, $\beta( M )$ should be scale-independent, so if the horizon-scale fluctuations have a Gaussian distribution with dispersion $\sigma_{\Hrm}$, one expects \cite{Carr:1975qj}
\begin{align}
	&\beta
		\approx {\rm Erfc}\!
				\left[
					\frac{\delta_{\rm c}}{\sqrt{2\.}\.\sigma_{\Hrm}}
				\right]
				,
				\label{eq:beta}
\end{align}
where Erfc is the complimentary error function and $\delta_{\crm}$ is the threshold for PBH formation. A simple analytic argument \cite{Carr:1975qj} suggests $\delta_{\crm} \approx \gamma$ but more precise arguments{\;---\;}both numerical \cite{Musco:2012au} and analytical \cite{Harada:2013epa}{\;---\;}suggest a somewhat larger value (see Ref.~\cite{Yoo:2018esr} for a recent investigation). Refs.~\cite{Hidalgo:2008mv, Polnarev:2012bi, Musco:2018rwt} discuss the dependence of the threshold on the shape of the desnity profile. Jedamzik {\it et al.} \cite{Jedamzik:2010dq} have studied matter-dominated scenarios associated with oscillating scalar fields during reheating, with possible saturations of PBH constraints being reported in Refs.~\cite{Gorbunov:2017fhq, Hidalgo:2017dfp}.

At one time it was argued that the primordial fluctuations would be {\it expected} to be scale-invariant \cite{1970PhRvD...1.2726H,1972MNRAS.160P...1Z} but this does not apply in the inflationary scenario. Nevertheless, one would still expect the above equations to apply if the PBHs to form from cosmic loops because the collapse probability is then scale-invariant. In this case Eq.~\eqref{eq:beta} is replaced by \cite{Hawking:1987bn}
\begin{align}
	\beta
		&\sim
				( G\.\mu )^{2 x - 4}
				\, ,
				\label{eq:loops}
\end{align}
where $\mu$ is the string tension and $x = L / s$ is the ratio of the string length to the correlation scale. One expects $2 < x < 4$ and requires $G\.\mu < 10^{-7}$ to avoid overproduction of PBHs. 

One usually assumes $0 < \gamma < 1$, corresponding to $2 < \alpha < 3$, in which case most of the density is in the smallest PBHs and the density of those larger than $M$ is 
\begin{align}
	\rho( M )
		&=
				\int^{ M_{\rm max} }_{ M } \d \tilde{ M }\;\.
				\tilde{ M }\.\frac{ \d n }{ \d \tilde{ M } }
		\propto
				M^{2-\alpha}
				\quad
				(
					M_{\rm min}
					<
					M
					<
					M_{\rm max}
				)
				\; ,
\end{align} 
where $M_{\rm max}$ and $M_{\rm min}$ are the upper and lower cut-offs for the mass function. 
If we assume that the PBHs contain a fraction $f_{\rm DM}$ of the dark matter, this implies
that the fraction of the dark matter (DM) in PBHs of mass larger than $M$ is 
\begin{align}
	f( M )
		&\equiv
				\frac{ \rho( M ) }{ \rho_{\rm DM} }
		\approx
				f_{\rm DM}
				\left(
					\frac{ M_{\rm DM} }{ M }
				\right)^{\!\alpha - 2}
				\quad
				(
					M_{\rm min}
					<
					M
					<
					M_{\rm max}
				)
				\; ,
				\label{eq:dark}
\end{align}
where $M_{\rm DM} \approx M_{\rm min}$ is the mass-scale which contains most of the dark matter. \footnote{Alternatively, one could define $f( M )$ as the fraction in PBHs in the mass interval ($M, 2\mspace{1mu}M$), which is smaller by a factor $1 - 2^{2 - \alpha}$.} In a radiation-dominated era, which is most likely, $\gamma = 1 / 3$ and the exponent in Eq.~\eqref{eq:dark} becomes $1 / 2$. The relationship between $f_{\rm DM}$ and the collapse fraction $\beta$ is then
\begin{align}
	f_{\rm DM}
		&\sim
				10^{8}\.\beta( M_{\rm DM} / M_{\odot})^{-1/2}
				\, .
				\label{eq:fbeta}
\end{align}
The above analysis also applies if the PBHs form from the collapse of cosmic strings but with $\beta$ given by Eq.~\eqref{eq:loops}.

%%%%%%%%%%%%%%%%%%%%%%%%%%%%%%%%%%%%%%%%%%%%%%%%%%%%%%%%%
\subsection{Collapse in a Matter-Dominated Era}
\label{sec:Collapse-in-a-Matter--Dominated-Era-II}

PBHs form more easily if the Universe becomes pressureless (ie.~matter-dominated) for some period. For example, this may arise at a phase transition in which the mass is channeled into non-relativistic particles \cite{Khlopov:1980mg, 1982SvA....26..391P} or due to slow reheating after inflation \cite{Khlopov:1985jw, Carr:1994ar}. In a related context, Hidalgo {\it et al.} have recently studied \cite{Hidalgo:2017dfp} PBH formation in a dust-like scenario of an oscillating scalar field during an extended period of preheating. Since the value of $\alpha$ in the above analysis is $2$ for $\gamma = 0$, one might expect $\rho( M )$ to increases logarithmically with $M$. However, the analysis breaks down in this case because the Jeans length is much smaller than the particle horizon, so pressure is not the main inhibitor of collapse. Instead, collapse is prevented by deviations from spherical symmetry and the probability of PBH formation can be shown to be \cite{Khlopov:1980mg} 
\begin{align}
	\beta( M )
		&=
				0.02\,\sigma_{\Hrm}( M )^{5}
				\; .
\end{align}
This is in agreement with the recent analysis of Harada et al.~\cite{Harada:2016mhb} and leads to a mass function
\begin{align}
	\frac{\d n}{\d M}
		&\propto
				M^{-2}\.\sigma_{\Hrm}( M )^{5}
				\; .
\end{align}
$\beta( M )$ is still small for $\delta_{\Hrm}( M ) \ll 1$ but much larger than the exponentially suppressed fraction in the radiation-dominated case. If the matter-dominated phase extends from $t_{1}$ to $t_{2}$, PBH formation is enhanced over the mass range 
\begin{align}
	M_{\rm min}
		&\sim
				M_{\Hrm}( t_{1} ) < M < M_{\rm max}
		\sim
				M_{\Hrm}(t_{2})\.\sigma_{\Hrm}
				(
					M_{\rm max}
				)^{3 / 2}
				\; .
\end{align} 
The lower limit is the horizon mass at the start of matter-dominance and the upper limit is the horizon mass at the epoch when the region which binds at the end of matter-dominance enters the horizon. This scenario has recently been studied in Ref.~\cite{Carr:2017edp}.

Since the primordial fluctuations must be approximately scale-invariant (even in the inflationary scenario), $\beta( M )$ is nearly constant, so Eq.~\eqref{eq:spectrum} applies with $\alpha \approx 2$. Thus the mass function is uniquely determined by the values of $t_{1}$ and $t_{2}$. Also $f( M )$ should only have a weak dependence on $M$ (logarithmic if $\beta$ is {\it exactly} constant).

%%%%%%%%%%%%%%%%%%%%%%%%%%%%%%%%%%%%%%%%%%%%%%%%%%%%%%%%%
\subsection{Collapse from Inflationary Fluctuations}
\label{sec:Collapse-from-Inflationary-Fluctuations-II}

If the fluctuations generated by inflation have a blue spectrum (ie.~decrease with increasing scale) and the PBHs form from the high-$\sigma$ tail of the fluctuation distribution, then the exponential factor in Eq.~\eqref{eq:beta} might suggest that the PBH mass function should have an exponential upper cut-off at the horizon mass when inflation ends \cite{Carr:1994ar}. This corresponds to the reheat time $t_{\rm R}$, which the cosmic microwave background (CMB) quadrupole anisotropy requires to exceed $10^{-35}\.$s. In this case, $f( M )$ should fall off exponentially above the reheat horizon mass. However, a more careful analysis gives a different result. If the fluctuations result from a smooth symmetric peak in the inflationary power spectrum \cite{GarciaBellido:1996qt}, the PBH mass function should have the lognormal form
\begin{align}
	\frac{\d n}{\d M}
		&\propto
				\frac{ 1 }{ M^{2} }\.
				\exp\!
				\left[
					-
					\frac{
						(
							\log M
							-
							\log M_{\crm}
						)^{2} }
					{ 2\.\sigma_{\Hrm}^{2} }
				\right]
				.
				\label{eq:mf}
\end{align}
This was first suggested by Dolgov \& Silk \cite{Dolgov:1992pu} and later by Clesse \& Garc{\'i}a-Bellido \cite{Clesse:2015wea}. It has been demonstrated both numerically \cite{Green:2016xgy} and analytically \cite{Kannike:2017bxn} for the case in which the slow-roll approximation holds. It is therefore representative of a large class of inflationary scenarios, including the axion-curvaton and running-mass inflation models considered by K{\"u}hnel et al.~\cite{Kuhnel:2015vtw}. Equation \eqref{eq:mf} implies that the mass function is symmetric about its peak at $M_{\crm}$ and described by two parameters: the mass-scale $M_{\crm}$ itself and the width of the distribution $\sigma$. The integrated mass function is
\begin{align}
	f( M )
		&=
				\int_{M}\d \tilde{M}\;
				\tilde{M}\.\frac{\d n}{\d \tilde{M}}
		\approx
				{\rm Erfc}\!
				\left(
					\ln \frac{M}{ \sigma_{\Hrm} }
				\right)
				.
				\label{eq:f(M)-Collapse-from-Inflation}
\end{align}
However, not all inflationary scenarios produce the mass function \eqref{eq:mf}. Inomata et al.~\cite{Inomata:2016rbd} propose a scenario which combines a broad mass function at low $M$ (to explain the dark matter) with a sharp one at high $m$ (to explain the LIGO events).

%%%%%%%%%%%%%%%%%%%%%%%%%%%%%%%%%%%%%%%%%%%%%%%%%%%%%%%%%
\subsection{Critical Collapse}
\label{sec:Critical-Collapse-II}

It is well known that black-hole formation is associated with critical phenomena \cite{Choptuik:1992jv} and the application of this to PBH formation has been studied by various authors \cite{Koike:1995jm, Niemeyer:1997mt, Evans:1994pj, Kuhnel:2015vtw}. The conclusion is that the mass function has an upper cut-off at around the horizon mass but there is also a low-mass tail \cite{Yokoyama:1998qw}. If we assume for simplicity that the density fluctuations have a monochromatic power spectrum on some mass scale $K$ and identify the amplitude of the density fluctuation when that scale crosses the horizon, $\delta$, as the control parameter, then the black-hole mass is \cite{Choptuik:1992jv} 
\begin{align}
	M
		&=
				K\.
				\big(
					\delta
					-
					\delta_{\crm}
				\big)^{\mspace{-1mu}\eta}
				\; .
				\label{eq:2}
\end{align}
Here $K$ can be identified with a mass $M_{\frm}$ of order the particle horizon mass, $\delta_{\crm}$ is the critical fluctuation required for PBH formation and the exponent $\eta$ has a universal value for a given equation of state. For $\gamma = 1 / 3$ the usual situation), one has $\delta_{\crm} \approx 0.4$ and $\eta \approx 0.35$. Although the scaling relation \eqref{eq:2} is expected to be valid only in the immediate neighborhood of $\delta_{\crm}$, most black holes should form from fluctuations with this value because the probability distribution function declines exponentially beyond $\delta = \delta_{\crm}$ if the fluctuations are blue. Hence it is sensible to calculate the expected PBH mass function using Eq.~\eqref{eq:2}. 

This allows us to estimate the mass function independently of the form of the PDF of the primordial density fluctuations. A detailed calculation gives the mass function \cite{Yokoyama:1998qw}
\begin{align}
	\frac{\d n}{\d M}
		&\propto
				\!
				\left(
					\frac{ M }{\xi\,M_{\frm}}
				\right)^{\!1 / \eta - 1}\,
				\exp\!
				\left[
					-
					(
						1
						-
						\eta
					)\!
					\left(
						\frac{ M }{\eta\,M_{\frm}}
					\right)^{\!\!1 / \eta}\.
				\right]
				,
				\label{eq:initialMF}
\end{align}
where
\begin{align}
	\xi
		&\equiv
				\!
				\left(
					\frac{ 1 - \eta }{ s }
				\right)^{\!\eta}
				\; ,
	\quad
	s
		=
				\delta_{\crm} / \sigma
				\; ,
	\quad
	M_{\frm}
		=
				K
				\, ,
\end{align}
and $\sigma$ is the dispersion of $\delta$. For $\eta = 0.35$, this gives
\begin{align}
	\frac{\d n}{\d m}
		&\propto
				M^{1.85}\.
				\exp\!
				\Bigg[
					-
					s\mspace{-1mu}
					\left(
						\frac{ M }{ M_{\frm} }
					\right)^{\!2.85}
				\Bigg]
				\, .
				\label{eq:critmf}
\end{align}
The function $f( M )$ would have a similar form but with an exponent of $3.85$ in the first term, so the PBH density is concentrated around the mass $K$.

The above analysis depends on the assumption that the power spectrum of the primordial fluctuations is monochromatic. As shown by K{\"u}hnel et al.~\cite{Kuhnel:2015vtw} for a variety of inflationary models, when a realistic model for the power spectrum is used, the inclusion of critical collapse can lead to a significant shift, lowering and broadening of the PBH mass spectra{\;---\;}sometimes by several orders of magnitude. \footnote{A further assumption is the sphericity of the collapse; deviating from this can induce large uncertainties (see Refs.~\cite{Kuhnel:2016exn, Akrami:2016vrq, Sato-Polito:2019hws}).}

%%%%%%%%%%%%%%%%%%%%%%%%%%%%%%%%%%%%%%%%%%%%%%%%%%%%%%%%%
\subsection{QCD Phase Transition}
\label{sec:QCD-Phase-Transition-II}

At one stage it was thought that the QCD phase transition might be first-order. This would mean that the quark-gluon plasma and hadron phases could coexist, with the cosmic expansion proceeding at constant temperature by converting the quark-gluon plasma to hadrons. The sound speed would then vanish and the effective pressure would be reduced, significantly lowering the threshold $\delta_{\crm}$ for collapse. PBH production during a first-order QCD phase transitions was first suggested by Crawford \& Schramm \cite{Crawford:1982yz} and later revisited by Jedamzik \cite{Jedamzik:1996mr}. The amplification of density perturbations due to the vanishing of the speed of sound during the QCD transition was also considered by Schmid and colleagues \cite{Schmid:1998mx, Widerin:1998my}, while Cardall \& Fuller developed a semi-analytic approach for PBH production during the transition \cite{Cardall:1998ne}. It is now thought unlikely that the QCD transition is 1st order but one still expects some softening in the equation of state. Recently Byrnes et al.~\cite{Byrnes:2018clq} have discussed how this softening{\;---\;}when combined with critical phenomena and the exponential sensitivity of $\beta( M )$ to the equation of state{\;---\;}could produce a significant change in the mass function. We use their numerical results in our later analysis.

%%%%%%%%%%%%%%%%%%%%%%%%%%%%%%%%%%%%%%%%%%%%%%%%%%%%%%%%%
\section{Vacua}
\label{sec:Vacua-2}

\noindent We consider a flat Friedmann-Lema{\^i}tre-Robertson-Walker universe with metric 
\begin{align}
 	\d s^{2}
		&=
				g_{\mu \nu}\.\d x^{\mu} \d x^{\nu}
		=
				a^{2}( \eta )
				\big(
					\d\eta^{2}
					-
					\d\vec{x}^{2}
				\big)
				\, ,
				\label{eq:line-element}
\end{align}
where the conformal time $\eta$ runs from $- \infty$ to $0$ and we put $c = 1$. For power-law inflation, the scale factor is
\begin{subequations}
\begin{align}
	a( \eta )
		&\propto
				| \eta |^{\frac{ p }{ 1 - p }}
				\; 
				\label{eq:scale-factor-power-law}
\end{align}
with $p > 1$. The limit $p \to \infty$ corresponds to the de Sitter solution,
\begin{align}
	a( \eta )
		&\propto
				-
				1 / \eta
				\; ,
				\label{eq:scale-factor-de-Sitter}
\end{align}
\end{subequations}
while $p = 0$ corresponds to Minkowski space. If we consider a minimally coupled scalar field $\phi$ with a potential $\Vrm( \phi )$, the action is
\begin{align}
	\Scal
		&=
				\int \d^4 x\;\sqrt{|\det( g )|}
				\left[
					\frac{ M_{\rm Pl} }{ 2 }\.\Rrm
					-
					\frac{1}{2}(\partial \phi)^{2}
					-
					\Vrm( \phi )
				\right]
				,
				\label{eq:Action}
\end{align}
where $\Rrm$ is the Ricci scalar and $M_{\rm Pl}^{2} \equiv ( 8 \pi G )^{-1/2}$ is the reduced Planck mass (set to $1$ in what follows). 

We perturb the metric and scalar field as follows:
\begin{subequations}
\begin{align}
	g_{\mu\nu}\.\d x^{\mu} \d x^{\nu}
		&\ral
				\big(
					g_{\mu\nu}
					+
					\delta g_{\mu\nu}
				\big)
				\d x^{\mu}\.
				\d x^{\nu}
		=
				a^{2}( \eta )\.
				\Big(					
					\big[
						1
						+
						2\Phi( \eta,\vec{x} )
					\big]\.
					\d\eta^{2}
					-
					\big[
						1
						-
						2\Phi( \eta,\vec{x} )
					\big]\.
					\d \vec{x}^{\mspace{2mu}2}\.
				\Big)
				\,
				,
				\label{eq:perturbed-metric}
				\\[2mm]
	\phi(\eta)
		&\ral
				\phi(\eta)
				+
				\delta\phi
				\big(
					\eta, \vec{x}\mspace{1mu}
				\big)
				\; ,
				\label{eq:perturbed-scalar-field}
\end{align}
\end{subequations}
where we define the Hubble and other parameters by
\begin{subequations}
\begin{align}
	H( \eta )
		&\equiv
				a'( \eta ) / a^{2}( \eta )
				\; ,
				\\[1.8mm]
	z( \eta )
		&\equiv
				\frac{ \phi'( \eta ) }{ H( \eta ) }
				\label{eq:z}
				\; ,
				\\
	v\big( \eta, \vec{x}\mspace{1mu} \big)
		&\equiv
				a( \eta )\.\delta \phi
				\big(
					\eta, \vec{x}\mspace{1mu}
				\big)
				+
				\frac{ {\phi}'( \eta ) }{ H( \eta ) }\,
				\Phi
				\big(
					\eta, \vec{x}\mspace{1mu}
				\big)
				\, ,
				\label{eq:v}
\end{align}
\end{subequations}
with a prime denoting $\d / \d\eta$. The second-order action of the perturbation $v$ then becomes \cite{Mukhanov:1990me}
\begin{align}
	\Scal^{(2)}
		&=
				\frac{ 1 }{ 2 }
				\int \d^{4}x\.
				\left[
					v'\big( \eta, \vec{x}\mspace{1mu} \big)^{2}
					-
					\Big[
						\nabla v\big( \eta, \vec{x}\mspace{1mu} \big)
					\Big]^{2}
					+
					\frac{ z''( \eta ) }{ z( \eta ) }
					v^{2}\big( \eta, \vec{x}\mspace{1mu} \big)
				\right]
				, 
				\label{eq:v-action}
\end{align}
which yields the equation of motion
\begin{align}
	v''\big( \eta, \vec{x}\mspace{1mu} \big)
	-
	\left[	
		\nabla^{2}
		+
		\frac{z''( \eta )}{z( \eta )}
	\right]
	v\big( \eta, \vec{x}\mspace{1mu} \big)
		&=
				0
				\; .
				\label{eq:EOM-v}
\end{align}
The classical variable $v$ can be quantised in the standard way. This turns it into an operator $\hat{v}$, which can be decomposed into Fourier modes: 
\begin{align}
	\hat{v}
	\big(
		\eta, \vec{x}\mspace{1mu}
	\big)
		&=
			\frac{1}{\sqrt{2\.}}
			\int\frac{\d \vec{k} }
			{ (2 \pi)^{3 / 2} }
			\left[
				\.v_{k}( \eta )\.
				e^{i \vec{k}\cdot\vec{x}}\.
				\hat{a}_{k}
				+
				v_{k}^{*}( \eta )\.
				e^{- i \vec{k}\cdot\vec{x}}\.
				\hat{a}^{\dag}_{k}\.
			\right]
			.
			\label{eq:mode-expansion}
\end{align}

If the mode functions $v_{k}$ satisfy the normalisation condition
\begin{align}
	v'_{k}( \eta )\.v^{*}_{k}( \eta )
	-
	v'^{*}_{k}( \eta )\.v^{}_{k}( \eta )
		&=
				-
				2\.i
				\; ,
				\label{eq:normalisation-condition}
\end{align}
the $\hat{a}_{k}$ can be interpreted as annihilation operators,
\begin{align}
	\Big[
		\hat{a}_{k}
		,\.
		\hat{a}_{k'}^{\dag}
	\Big]
		&=
				\delta
				\big(
					\vec{k}
					-
					\vec{k'}
				\big)
				\, ,
\end{align}
with 
\begin{align}
	\hat{a}_{k} | \Omega \rangle
		&=
				0
\end{align}
defining a class of vacua $| \Omega \rangle$. However, this does not uniquely determine the vacua. In order to do so, the initial conditions for the equation of motion \eqref{eq:EOM-v} have to be specified. To this end, we denote the dimensionless ratio of the physical extent of the perturbation,
\begin{align}
	\lambda( \eta )
		&\equiv
				\frac{ a( \eta ) }{ k }
				\; ,
				\label{eq:Xi}
\end{align}
and the Hubble radius,
\begin{align}
	r_{\rm H}( \eta )
		&\equiv
				\frac{ 1 }{ H( \eta ) }
				\; ,
				\label{eq:rH} 
\end{align}
by
\begin{align}
	\theta( \eta )
		&\equiv
				\frac{ \lambda( \eta ) }{ r_{\rm H}( \eta ) }
		=
				\frac{ p }{ 1 - p }\.
				\frac{ 1 }{ k\.\eta }
				\; .
				\label{eq:theta}
\end{align}
This is positive for $p > 1$ and $\eta < 0$. We denote by $\eta_{0}$ the conformal time at which the initial conditions are imposed and let $\theta_{0} \equiv \theta( \eta_{0} )$.

The standard choice of the initial conditions is the one introduced by Bunch and Davies \cite{Bunch:1978yq}. This relies on the fact that the effect of inflation should not matter far below the Hubble scale, so that space-time is essentially flat. One should therefore recover Minkowski space in the limit $\eta_{0} \rightarrow - \infty$, with the mode functions assuming the form
\begin{subequations}
\begin{align}
	v_{k}
		&=
				\frac{ 1 }{ \sqrt{k\.} }
				\; ,
				\label{eq:vk-Minkowkski}
				\\[1.5mm]
	v'_{k}
		&=
				-
				i\.\sqrt{k\.}
				\; .
				\label{eq:vk'-Minkowkski}
\end{align}
\end{subequations}
However, the physical length-scale $\lambda$ tends to $0$ in this limit, so conventional physics cannot be applied. One therefore needs to set the initial conditions at a time when a given mode is much larger than the Planck scale. It is convenient to choose this mode to be well below the Hubble radius, where the cosmological expansion is negligible. One may then take $\eta_{0}$ to be the time when the physical length of the mode equals some fixed scale $L \geq L_{\rm Pl}$, so that
\begin{align}
	\frac{ a( \eta_{0} ) }{ k }
		&=
				L
				\; .
				\label{eq:Xi}
\end{align}

At early times, such that $\theta_{0} \ll 1$, one can expand the mode functions as \cite{ArmendarizPicon:2003gd}
\begin{subequations}
\begin{align}
	v_{k}( \eta_{0})
		&=
				\frac{ 1 }{ \sqrt{k\.} }
 				\left[
					1
					+
					\left(
						\frac{ X + Y }{ 2 }
					\right)
					\theta_{0}
					+
					\Ocal
					\big(
						\theta_{0}^{2}
					\big)
				\right]
				,
				\label{eq:vk-general}
				\\[2mm]
	v_{k}'( \eta_{0} )
		&=
				-
				i\.\sqrt{k\.}
				\left[
					1
					+
					\left(
						\frac{ Y - X }{ 2 }
					\right)
					\theta_{0}
					+
					\Ocal
					\big(
						\theta_{0}^{2}
					\big)
				\right]
				,
				\label{eq:vkprime-general}
\end{align}
\end{subequations}
with $X,\.Y \in \Cbb$ and $\mathfrak{Re}( Y ) = 0$ due to the normalisation condition \eqref{eq:normalisation-condition}. The standard (Bunch-Davies) scenario is recovered for $X = 0$ and $Y = i\.( 1 - 2 p ) / ( 1 - p )$. With the choice \ref{eq:vk-general}{\color{midblue},b}), the solution of Eq.~\eqref{eq:EOM-v} is \cite{ArmendarizPicon:2003gd}
\begin{align}
	v_{k}( \eta )
		&=
				| \eta |^{1 / 2}
				\big[
					\.A^{}_{k}\.\Hrm^{}_{\nu}( | k \eta | )
 					+
					B^{}_{k}\.\Hrm^{*}_{\nu}( | k \eta | )\mspace{1mu}
				\big]
				\, ,
 				\label{eq:general-solution}
\end{align}
where the $\Hrm_{\nu}$ are Hankel functions and
\begin{align}
	\nu
		&\equiv
				\frac{ 3 }{ 2 }
				+
				\frac{ 1 }{ p - 1 }
				\; .
				\label{eq:nu}
\end{align}
The time-independent complex functions $A_{k}$ and $B_{k}$ become \cite{ArmendarizPicon:2003gd}
\begin{subequations}
\begin{align}
	A_{k}
		&=
				\sqrt{ \frac{ \pi }{ 2 } }
				\left[
					1
					+
					Y\,
					\frac{ \theta_{0} }{ 2 }
					-
					\frac{ i }{ 2 }\,
					\frac{ 1 - 2\.p }{ 1 - p }\.\theta_{0}
				\right]
				e^{i \varphi}
				+
				\Ocal
				\big(
					\theta_{0}^{2}
				\big)
				\; ,
				\label{eq:Ak}
				\\[1.5mm]
	B_{k}
		&=
				\sqrt{ \frac{ \pi }{ 2 }\.}\,
				X\,
				\frac{ \theta_{0} }{ 2 }\.
				e^{-i\varphi}
				+
				\Ocal
				\big(
					\theta_{0}^{2}
				\big)
				\label{eq:Bk}
				\; ,
				\\[1.5mm]
	\varphi
		&=
				\frac{ p }{ 1 - p }\,\frac{ 1 }{ \theta_{0} }
				\pm \frac{ \pi }{ 2 }
				\left(
					\frac{ 1 - 2\.p }{ 1 - p }
				\right)
				+
				\Ocal
				\big(
					\theta_{0}
				\big)
				\; .
				\label{eq:varphi}
\end{align}
\end{subequations}
The upper sign in Eq.~\eqref{eq:varphi} holds for $p > 1$ and the lower sign for $p < 1$. Using the curvature perturbation in the comoving gauge (see Refs.~\cite{Mukhanov:1990me, Malik:2008im} for the conventional definitions),
\begin{align}
	\hat{\zeta}( \eta, \vec{x} )
		&\equiv
				\sqrt{4\pi\.p\.}\;
				\frac{ \hat{v}( \eta, \vec{x} ) }{ a( \eta ) }
				\; ,
				\label{eq:curvature-perturbation}
\end{align}
the associated power spectrum $\Pcal_{k}( \eta )$ can be defined via
\begin{align}
	\Big\langle \Omega \Big|\.
		\hat{\zeta}^{\dag}
		\big(
			\eta, \vec{x}\mspace{1mu}
		\big)\.
		\hat{\zeta}
		\big(
			\eta, \vec{x} + \vec{r}\.
		\big)
	\Big| \Omega \Big\rangle
		&=
				\int\frac{ \d k }{ k }\;
				\frac{ \sin\!\big( k\.| \vec{r}\. | \big) }
				{ k\.| \vec{r}\. | }\,				\frac{ p }{ \pi }\.
				\frac{k^{3} \big| v_{k}( \eta ) \big|^{2} }{ a^{2}( \eta ) }
		\equiv
				\int\frac{ \d k }{ k }\;
				\frac{ \sin\!\big( k\.| \vec{r}\. | \big) }
				{ k\.| \vec{r}\. | }\,\Pcal^{}_{k}( \eta )
				\; .
				\label{eq:power-def}
\end{align}
Using the mode functions (\ref{eq:Ak}{\color{midblue},b}) yields \cite{ArmendarizPicon:2003gd} 
\begin{align}
	\Pcal^{}_{k}( \eta )
		&=
				\Pcal^{(0)}_{k}\.
				\big[
					1
					-
					\mathfrak{Re}( X )\,
					\theta_{0}
					\cos
					\Theta
					-
					\mathfrak{Im}( X )\,\theta_{0}
					\sin
					\Theta
				\big]
				\, ,
				\label{eq:P}
\end{align}
with
\begin{align}
	\Theta
		&\equiv
				\frac{ 2\.p }
				{ ( 1 - p )\.\theta_{0} }
				+
				\frac{ \pi\,( 1 - 2\.p )}
				{ 1 - p }
				\; 
				\label{eq:Theta-k}
\end{align}
and $\Pcal^{(0)}_{k}$ being the standard power spectrum. 

For the Bunch-Davies vacuum, given by Eq.~(\ref{eq:vk-Minkowkski}{\color{midblue},b}), we have
\begin{align}
	\Pcal^{(0)}_{k}
		&=
				\widetilde{\Pcal}_{k_{*}}
				\left(
					\frac{ k }{k_{*}}
				\right)^{\mspace{-4mu}n_{\rm s} - 1}
				\; ,
				\label{eq:P0}
\end{align}
where $k_{*}$ is some pivot scale. The spectral index $n_{\rm s}$ for power-law inflation is
\begin{align}
	n_{\rm s}
		&=
				1
				-
				\frac{ 2 }{ p - 1 }
				\; 
				\label{eq:ns}
\end{align}
and the prefactor becomes \cite{ArmendarizPicon:2003gd}
\begin{align}
	\widetilde{\Pcal}^{}_{k_{*}}
		&=
				\frac{ p\,|\Gamma(\nu)|^{2} }{ \pi^{2} } 
				\left(
					\frac{2\.( p - 1 ) }{ p }
				\right)^{\mspace{-4mu}n_{\rm s} - 1}
				( H_{*}\.L )^{n_{\rm s} - 1}
				H_{*}^{2}
				\; ,
				\label{eq:Pstar}
\end{align}
where $H_{*}$ denotes the value of the Hubble parameter when $k_{*}$ equals $1 / L$.

This modulation of the power spectrum is a general finding. While the key result Eq.~\eqref{eq:P} has been derived for power-law inflation with different choices for the vacua, it is independent of the specific choice of the power-law exponent. For instance, Bergstr{\"o}m and Danielson~\cite{Bergstrom:2002yd} have used de Sitter inflation and arrived at
\vs{-1mm}
\begin{align}
	\frac{ \Pcal^{}_{k} }{ \Pcal^{(0)}_{k} }
		&=
				1
				+
				\theta_{0}
				\sin\!
				\left(
					\frac{ 2 }{ \theta_{0} }
				\right)
				,
				\label{eq:P-de-Sitter}
\end{align}
which is obtained from Eq.~\eqref{eq:P} by choosing $X \equiv - i \equiv - Y$ and applying the limit $p \ra \infty$. For sufficiently small values of the slow-roll parameter, $\varepsilon \equiv ( \Vrm' / \Vrm )^{2} / 2$, Eq.~\eqref{eq:P-de-Sitter} can be recast as
\begin{align}
	\frac{ \Pcal_{k} }{ \Pcal_{k}^{(0)} }
		&\simeq
				1
				+
				\lambda\.
				\sin\!
				\left[
					\.\nu\.
					\log\!
					\left(
						\frac{ k }{ k_{*} }
					\right)
				\right]
				,
				\label{eq:P-de-Sitter-parametrised-2}
\end{align}
where $\nu \equiv 2\.\varepsilon / \lambda$. The parameter $\lambda$ gives the amplitude of the oscillatory onset, $\nu$ determines its frequency and the scale $k_{*}$ defines its phase. We will use this general parametrisation in Sec.~\ref{sec:PBH-Scenarios} for a variety of different power spectra underlying the PBH formation scenarios outlined in Sec.~\ref{sec:PBH-Formation}.

%%%%%%%%%%%%%%%%%%%%%%%%%%%%%%%%%%%%%%%%%%%%%%%%%%%%%%%%%
\section{Resonance Effects}
\label{sec:Resonance-Effects}

\noindent Cai et al.~\cite{Cai:2018tuh} have discussed a new type of resonance effect which leads to prolific PBH formation. This arises because the sound-speed can oscillate in some inflationary scenarios, leading to parametric amplification of the curvature perturbation $\zeta$ and hence a significant peak in the power-spectrum of the density perturbations on some critical scale. In a single-field model with a non-canonical kinetic term, the sound-speed $c_{\rm s}$ can deviate from $1$ according to
\begin{align}
	c_{\rm s}^{2}
		&=
				1
				-
				2\.\xi\mspace{1mu}
				\big[
					1
					-
					\cos\!
					\left(
						2\.k_{\rm s}\.\eta
					\right)\mspace{-2mu}
				\big]
				\, ,
				\label{eq:cs}
\end{align}
where $\xi$ and $k_{\rm s}$ are the oscillation amplitude and frequency. This can happen in UV-complete theories, such as D-brane dynamics in string theory. The perturbations are described by a Mathieu equation and this corresponds to an instability for some range of $k$, leading to a transfer of energy from the inflaton into other particles during preheating.

The resonances are in narrow bands around the harmonic frequencies $k = n\.k_{\rm s}$, with $n = 1$ dominating. The amplification stops when the scale exits the Hubble horizon at $\eta_{\rm p} = - 1 / k_{\rm p}$ and the resulting power spectrum is
\begin{align}
	\Pcal^{\zeta}_{k}
		&\approx
				A_{s}
				\left(
					\frac{ k }{ k_{\rm p} }
				\right)^{\!n_{\rm s} - 1}
				\left[
					1
					+
					\frac{ \xi\.k_{\rm p} }{ 2 }\.
					e^{- \xi\.k_{\rm p} \eta_{0}}
					\.\delta( k - k_{\rm p} )
				\right]
					,
\end{align}
with $k_{\rm p} \approx 0.05 \, {\rm Mpc}^{-1}$ being the pivot scale. The first term corresponds to the usual (nearly scale-invariant) component. The second term is the resonance and arises from the interplay of the oscillation of the sound-speed and the cosmic expansion. One can easily get a peak of order unity. The corresponding PBH mass scale is
\begin{align}
	M
		&\approx
				\frac{ 4 \pi\.\gamma }
				{H
				\big[
					t_{\rm{exit}}( k_{M} )
				\big]}\,e^{\Delta N( k_{M} )} 
				\; , 
\end{align}
where $\gamma \approx 0.2$, $k_{M}$ is the mode whose physical scale at Hubble re-entry is the Schwarzschild radius of the black hole and $\Delta N( k_{M} )$ is the number of e-folds between the Hubble-exit of this mode and its re-entry. The fraction of the Universe collapsing into the PBHs and the current dark-matter fraction are given by Eqs.~\eqref{eq:beta} and \eqref{eq:f(M)-Collapse-from-Inflation}, respectively. Although most PBHs form at the first peak, a small number will also form at subsequent peaks.

We stress that this resonant effect is physically distinct from the vacuum effect discussed in the previous section. However, both effects produce spikes in the power spectrum, leading to spikes in the dark-matter fraction $f( M )$, so there is clearly a phenomenological connection. Whereas the main spike is chosen to be of order $0.1$ in the vacuum case, in order to produce an interesting number of PBHs, the main spike is larger in the resonant case{\;---\;}indeed the power spectrum peaks at order unity, so there is a danger of overproducing PBHs. On the other hand, the higher-harmonic spikes are much smaller in the resonant case, whereas{\;---\;}depending on the form of the original power spectrum{\;---\;}they can be comparable in the vacuum case. Note that the spikes in the $f( M )$ diagram of Cai et al.~correspond to different choices of the parameter $k_{*}$ and not different harmonics. Conceivably, since both effects are supposed to arise naturally, one could have a combination of the two effects.

%%%%%%%%%%%%%%%%%%%%%%%%%%%%%%%%%%%%%%%%%%%%%%%%%%%%%%%%%
\section{Application to PBH Scenarios}
\label{sec:PBH-Scenarios}

\noindent The previous two sections discussed how various effects give rise to a modification of the primordial power spectrum $\Pcal_{k}$. We now apply these results to the various models for PBH formation discussed in Sec.~\ref{sec:PBH-Formation}. We focus on the oscillations originating from non-Bunch-Davies vacua (see Sec.~\ref{sec:Vacua-2} and emphasise the effect on the current PBH mass spectra. We stress that the class of allowed vacua discussed in Sec.~\ref{sec:Vacua-2} is very large, since the permitted values of $\Lambda$ span many orders of magnitude and there are different possible choices for the frequency, amplitude and phase. For all scenarios addressed below, we use Eq.~\eqref{eq:P-de-Sitter-parametrised-2} with $\lambda = 0.05$ and $\nu = 8$ as an example.

We will first focus on inflationary scenarios which allow PBH formation. We apply the critical collapse model and discuss how the oscillations induced in the power spectrum modify the mass spectra. Specifically, we will discuss running-mass inflation (Sec.~\ref{sec:Running-Mass}), hybrid inflation (Sec.~\ref{sec:Hybrid-Inflation}), axion-like curvaton inflation (Sec.~\ref{sec:Axion-Curvaton}) and the QCD phase transition (Sec.~\ref{sec:QCD-Phase-Transition}), these models being discussed in detail in Ref.~\cite{Kuhnel:2015vtw}. Although we will use Eq.~\eqref{eq:critmf} rather than Eq.~\eqref{eq:mf}, in practice there is little difference between them. We then discuss how scale-invariant spectra (Sec.~\ref{sec:Scale-Invariant}) are modified by the results of Sec.~\ref{sec:Vacua-2}. We do not discuss the matter-dominated scenario.

As discussed above, the key ingredient for PBH production is the early generation of curvature 
perturbations larger than some threshold value $\zeta_{\crm}$, these then collapsing into black holes after horizon re-entry. The amplitude of the curvature-perturbation power spectrum at the pivot scale{\;---\;}$k_{\rm p} = 0.002\.{\rm Mpc}^{-1}$ for WMAP \cite{Komatsu:2010fb} or $k_{\rm p} = 0.005\.{\rm Mpc}^{-1}$ for Planck \cite{Planck:2013jfk, Ade:2015lrj}{\;---\;}is far too small to produce a notable PBH abundance, so one needs the power spectrum to become large at some early time. The models below (with the exception of the scale-invariant case) all have the feature that their power spectra have an increased amplitude (ie.~a bump) at some small scale (high $k$).\\[-8mm]

%%%%%%%%%%%%%%%%%%%%%%%%%%%%%%%%%%%%%%%%%%%%%%%%%%%%%%%%%
\subsection{Running Mass}
\label{sec:Running-Mass}

\noindent PBH formation in the running-mass model \cite{Stewart:1996ey, Stewart:1997wg} has been intensively studied in Refs.~\cite{Bugaev:2008gw, Drees:2011yz, Drees:2011hb, Drees:2012sz}. (See also Ref.~\cite{Leach:2000ea} for a discussion on constraints.) K{\"u}hnel et al.~\cite{Kuhnel:2015vtw} have investigated PBH production in this model with critical collapse and we follow their description below.

The simplest realisation corresponds to the inflationary potential
\begin{align}
	\Vrm( \phi )
		&=
				V_{0}
				+
				\frac{ 1 }{ 2 }\.m_{\phi}^{2}( \phi )\.\phi^{2}
				\, ,
				\label{eq:pot1}
\end{align}
with $V_{0}$ being constant. There exists a plethora of embeddings of this model in various frameworks, such as hybrid inflation \cite{Linde:1993cn}, which lead to different specific functions $m_{\phi}( \phi )$. These yield distinct expressions for the primordial power spectra, whose variance can be written in the general form \cite{Drees:2011hb}:
\begin{align}
	\big[ \sigma( k ) \big]^{2}
		&\simeq
				\frac{ 8 }{ 81 }\.\Pcal_{k_{\rm p}}
				\bigg(
					\frac{ k }{ k_{\rm p} }
				\bigg)^{\!\! n( k ) - 1}
				\Gamma\!
				\left(
					\frac{ n_{\srm}( k ) + 3 }{ 2 }
				\right)
				\label{eq:sigma-running-mass}
				,
\end{align}
where the spectral indices $n( k )$ and $n_{\srm}( k )$ are given by 
\begin{subequations}
\begin{align}
	n( k )
		&=
				n_{\srm}( k_{\rm p} )
				-
				\frac{ 1 }{ 2! }\.a\.\ln\!
				\bigg(
					\frac{ k }{ k_{\rm p} }
				\bigg)
				+
				\frac{ 1 }{ 3! }\.b\.\ln^{2}\!
				\bigg(
					\frac{ k }{ k_{\rm p} }
				\bigg)
				-
				\frac{ 1 }{ 4! }\.c\.\ln^{3}\!
				\bigg(
					\frac{ k }{ k_{\rm p} }
				\bigg)
				+
				\ldots
				\; ,
				\label{eq:n-running-mass}
				\displaybreak[1]
				\\[2mm]
	n_{\srm}( k )
		&=
				n_{\srm}( k_{\rm p} )
				-
				a\.\ln\!
				\bigg(
					\frac{ k }{ k_{\rm p} }
				\bigg)
				+
				\frac{1}{2}\.b\.\ln^{2}\!
				\bigg(
					\frac{ k }{ k_{\rm p} }
				\bigg)
				-
				\frac{1}{6}\.c\.\ln^{3}\!
				\bigg(
					\frac{ k }{ k_{\rm p} }
				\bigg)
				+
				\ldots
				\; ,
\end{align}
\end{subequations}
with real parameters $a$, $b$, and $c$. The $a$ terms are referred to as ``running'', the $b$ terms as ``running-of-running'', and the $c$ terms as``running-of-running-of-running''. The expansion is in principle infinite to account for any functional form of the running at any value of $k$. However, for definiteness, we consider a model that includes only the first three terms.\footnote{In order for the power spectrum (and hence $\beta$) to prevent an overproduction of small PBHs, one needs to go beyond the second order terms in the expansions (\ref{eq:n-running-mass}{\color{midblue},b}).} \footnote{Note that the spectral index and amplitude of the primordial power spectrum at the pivot scale $k_{\rm p} = 0.002\.{\rm Mpc}^{-1}$ have been measured to be $n_{\rm s}( k_{\rm p} ) \approx 0.96 < 1$ and $\Pcal_{k_{\rm p}} = \Ocal( 10^{-9} )$~\cite{Komatsu:2010fb, Planck:2013jfk, Ade:2015lrj}, so models without running certainly cannot produce an appreciable abundance of PBHs. Furthermore, since $a = 0.003 \pm 0.007 \ll 1$ \cite{Ade:2015lrj}, running alone is insufficient to increase the power spectrum at early times. Hence one needs to include at least a running-of-running term, this being only weakly constrained: $b \simeq 0.02 \pm 0.02$ \cite{Planck:2013jfk, Ade:2015lrj}).}

\begin{figure}
	\centering
	\includegraphics[scale=1,angle=0]{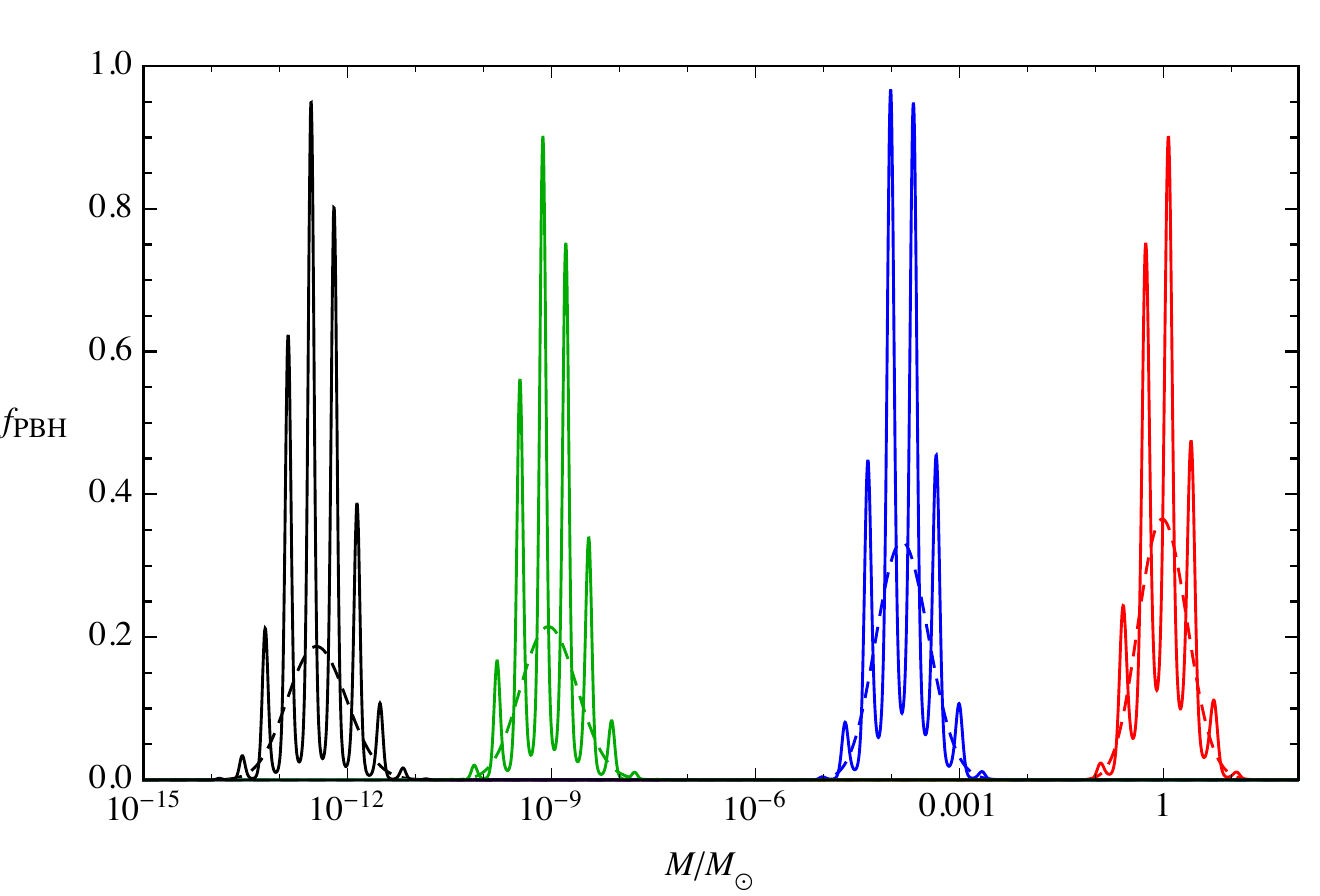}
	\caption{The present PBH dark-matter fraction $f_{\rm PBH}$ 
		implied by Eq.~\eqref{eq:sigma-running-mass} for the running-mass case, 
		including running, running-of-running and running-of-running-of-running, 
		as a function of $M / M_{\odot}$.
		A value $a = 0.011$ is used for all graphs, 
		while the individual ones have 
		(right to left) 
		$b = 0.02$, $c = - 0.0023405$ (red), 
		$b = 0.012$, $c = - 0.001224958$ (blue), 
		$b = 0.0065$, $c = - 0.0005728$ (green), 
		$b = 0.0045$, $c = - 0.0003675$ (black). 
		The dashed curves show the unmodulated results, 
		while the solid ones include a particular choice of non-Bunch-Davies vacua.
		}
	\label{fig:Running-Mass}
\end{figure}

Figure~\ref{fig:Running-Mass} shows the present PBH dark-matter fraction $f_{\rm PBH}$ as a function of mass for the four parameter choices indicated in the caption. Due to the exponential enhancement of the power spectrum, $f_{\rm PBH}$ exhibits pronounced oscillatory spikes. \footnote{Note that the values for $\lambda$ and $\nu$ in Fig.~\ref{fig:Running-Mass} apply for a particular vacuum choice. For a different choice, both the amplitude and frequency of the oscillations would be different.} The dark-matter fraction for each one is given by
\begin{align}
	f_{\rm PBH}
		&\sim
					\int\d\ln k\; {\rm erfc}\!
					\left(
						\frac{ \zeta_{c} }{ \sqrt{2\,}\sigma( k ) }
					\right)
					,
					\label{eq:fPBH-sim-int-erfc}
\end{align}
with integration over the values of $k$ encompassing the spike. In general, for a given set of constraints, these multimodal mass spectra allow a larger dark-matter fraction than the original ones. A more accurate investigation requires a careful integration of the extended spectra, as shown (for example) in Refs.~\cite{Kuhnel:2017pwq, Carr:2017jsz}.

%%%%%%%%%%%%%%%%%%%%%%%%%%%%%%%%%%%%%%%%%%%%%%%%%%%%%%%%%
\subsection{Hybrid Inflation}
\label{sec:Hybrid-Inflation}

\noindent Hybrid inflation is a two-field model, first introduced by Linde~\cite{Linde:1993cn}. It generically describes the situation in which inflation ends in a rapid rolling (``waterfall'') phase of one scalar field, triggered by the other one. PBH production in this model has been investigated by Clesse and Garc{\'i}a-Bellido \cite{Clesse:2015wea}. Using critical collapse, hybrid inflation has been reexamined in Ref.~\cite{Kuhnel:2015vtw}, which we follow below.

\begin{figure}
	\centering
	\includegraphics[scale=1,angle=0]{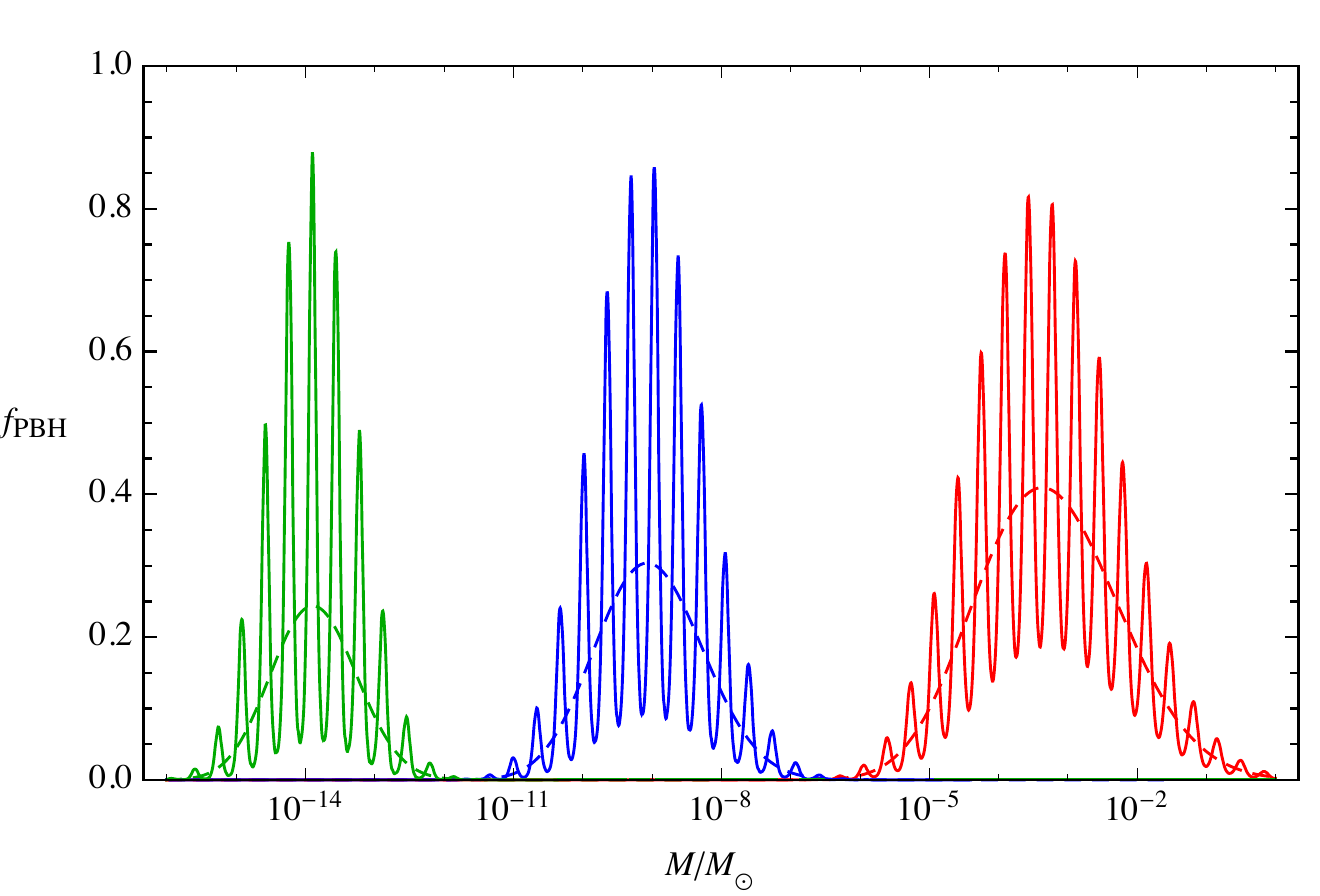}
	\caption{The present PBH dark-matter fraction $f_{\rm PBH}$ 
		for the hybrid-inflation case as a function of $M / M_{\odot}$.
		The parameter choices $\mu_{1} = 3 \times 10^{5}\.M_{\rm Pl}$, 
		$M = 0.1\.M_{\rm Pl}$ are used for all graphs,
		while the individual ones have (right to left) 
		$\phi_{\crm} = 0.125\.M_{\rm Pl}$, $\zeta_{\crm} = 1.43$ (red), 
		$\phi_{\crm} = 0.1\.M_{\rm Pl}$, $\zeta_{\crm} = 1.54$ (blue), and 
		$\phi_{\crm} = 0.1\.M_{\rm Pl}$, $\zeta_{\crm} = 1.63$ (green).
		The dashed curves show the unmodulated result, 
		while the solid lines include a particular choice of non-Bunch-Davies vacua.
		}
	\label{fig:Hybrid}
\end{figure}

The resulting power spectrum of the curvature perturbations $\zeta$ 
is \cite{Clesse:2015wea}
\begin{align}
		&\Pcal^{\zeta}_{k}
		\simeq
				\frac{ \Lambda\.M^{2}\.\mu_{1}\.\phi_{\crm} }
				{ 192\.\pi^{2}\.M_{\rm Pl}^{6}\.\kappa_{2}\.\psi_{k}^{2} }
				\; ,
				\label{eq:sigma-hybrid}
\end{align}
where
\begin{align}
	\psi_{k}
		&\equiv
				\psi_{0}\.\erm^{\kappa_{k}}
				\, ,
				\quad 
	\kappa_{k}
		\equiv
				\frac{ 4\.\phi_{\crm}\.\mu_{1} \xi_{k}^{2} }{ M^{2} }
				\, , 
				\quad
	\psi_{0}
		\equiv
				\left(
					\frac{ \Lambda\.\sqrt{2\.\phi_{\crm}\.\mu_{1}^{}\,}\.M }{ 96\.\pi^{3/2} }
				\right)^{\!\!1 / 2}
				\, .
\label{terms}
\end{align}
The durations of the two phases of inflation are given in terms of the number of $e$-foldings,
\begin{align}
	N_{1}
		&=
				\frac{ \sqrt{\kappa_{2}\.\phi_{\crm}\.\mu_{1}\,} M }{ 2 M_{\rm Pl}^{2} } 
				\; ,
				\quad
	N_{2}
		\simeq
				\frac{ M \sqrt{ \mu_{1}\.\phi_{\crm}\,} }
				{ 4\.M_{\rm Pl}^{2}\.\sqrt{\kappa_{2}\,} }
				\; .
\end{align}
and this specifies the quantity $\xi_{k}$ appearing in Eq.~\eqref{terms}:
\begin{align}
	\xi_{k}
		&\equiv
				-
				\frac{ M_{\rm Pl}^{2} \.( N_{1} + N_{2} - N_{k} ) }{ \mu_{1} \phi_{\crm} }
				\, .
\end{align}
For a given wavenumber $k$, exiting the Hubble radius $| N_{k} |$ $e$-foldings before the end of inflation, the associated PBH mass for horizon-scale collapse is (see Ref.~\cite{Clesse:2015wea} for details)
\begin{align}
	M_{k}
		&=
				\frac{ M_{\rm Pl} }{ \sqrt{ \Lambda / 3\,}\,\erm^{- 2 N_{k}} }
				\; .
\end{align}
Figure~\ref{fig:Hybrid} shows the present dark-matter fraction $f_{\rm PBH}$ as a function of mass for the three parameter choices indicated in the caption. As for the running-mass case, the modifications to the power spectrum \eqref{eq:sigma-hybrid} due to non-standard vacua yield pronounced oscillatory spikes in $f_{\rm PBH}$. These signatures have not been discussed before and characterise the vacuum choice. This could lead to more PBHs at the present epoch.

%%%%%%%%%%%%%%%%%%%%%%%%%%%%%%%%%%%%%%%%%%%%%%%%%%%%%%%%%
\subsection{Axion-Curvaton}
\label{sec:Axion-Curvaton}

\noindent In Refs.~\cite{Kohri:2012yw, Kawasaki:2012wr, Kuhnel:2015vtw} an axion-like curvaton model has been discussed in the context of PBH formation. See Refs.~\cite{Lyth:2001nq, Lyth:2002my} for a discussion of the original curvaton model and Ref.~\cite{Kasuya:2009up} for a discussion of the axion-like curvaton scenario. We follow Ref.~\cite{Kuhnel:2015vtw} in our description of these models below.
		
In this model the variance of the power spectrum due to the curvaton perturbations can be written as \cite{Kuhnel:2015vtw}:
\begin{align}
	\sigma^{2}( M_{H} )
		&=
				\frac{ 8 }{ 81 }\;{\widetilde\Pcal}^{\zeta}_{k}\!
				\left[
					\left(
						\frac{ M_{\frm} }{ M_{H} }
					\right)^{\!(n_{\chi}-1)/2}
					\gamma\mspace{-1mu}
					\left(
						\frac{ n_{\chi}-1 }{ 2 }, \frac{ M_{H} }{ M_{\frm} }
					\right)
					+
					E_{1}\!
					\left(
						\frac{ M_{H} }{ M_{f} }
					\right)
				\right]
				,
				\label{eq:sigma-axion-like}
\end{align}
for $M_{H} > M_{\rm f}$ and 
zero otherwise. Here
\begin{align}
	M_{\frm}
		&\approx
				10^{13 - 12 / ( n_{\chi} - 1 )}
				\left(
					\frac{ {\widetilde\Pcal}^{\zeta}_{k} }
					{ 2 \times 10^{-3} }
				\right)^{\!\!- 2 / ( n_{\chi} - 1 )}
				M_{\odot} 
\end{align}
and 
\begin{align}
	n_{\chi} - 1
		&=
				3
				-
				3\.\sqrt{ 1 - 4 / 9\.\lambda\,}
				\; ,
\end{align}
with $\lambda \in (1,\.9 / 4]$. Also 
\begin{align}
	\gamma( a, x )
		&\equiv
				\int_{0}^{x}\d t\;t^{a - 1}\erm^{-t}\, ,
				\quad
	E_{1}( x )
		\equiv
				\int_{x}^{\infty}\d t\;\frac{ \erm^{-t} }{ t }
				\; .
\end{align}
These being the lower incomplete gamma function and the exponential integral, respectively.

\begin{figure}
	\centering
	\includegraphics[scale=1,angle=0]{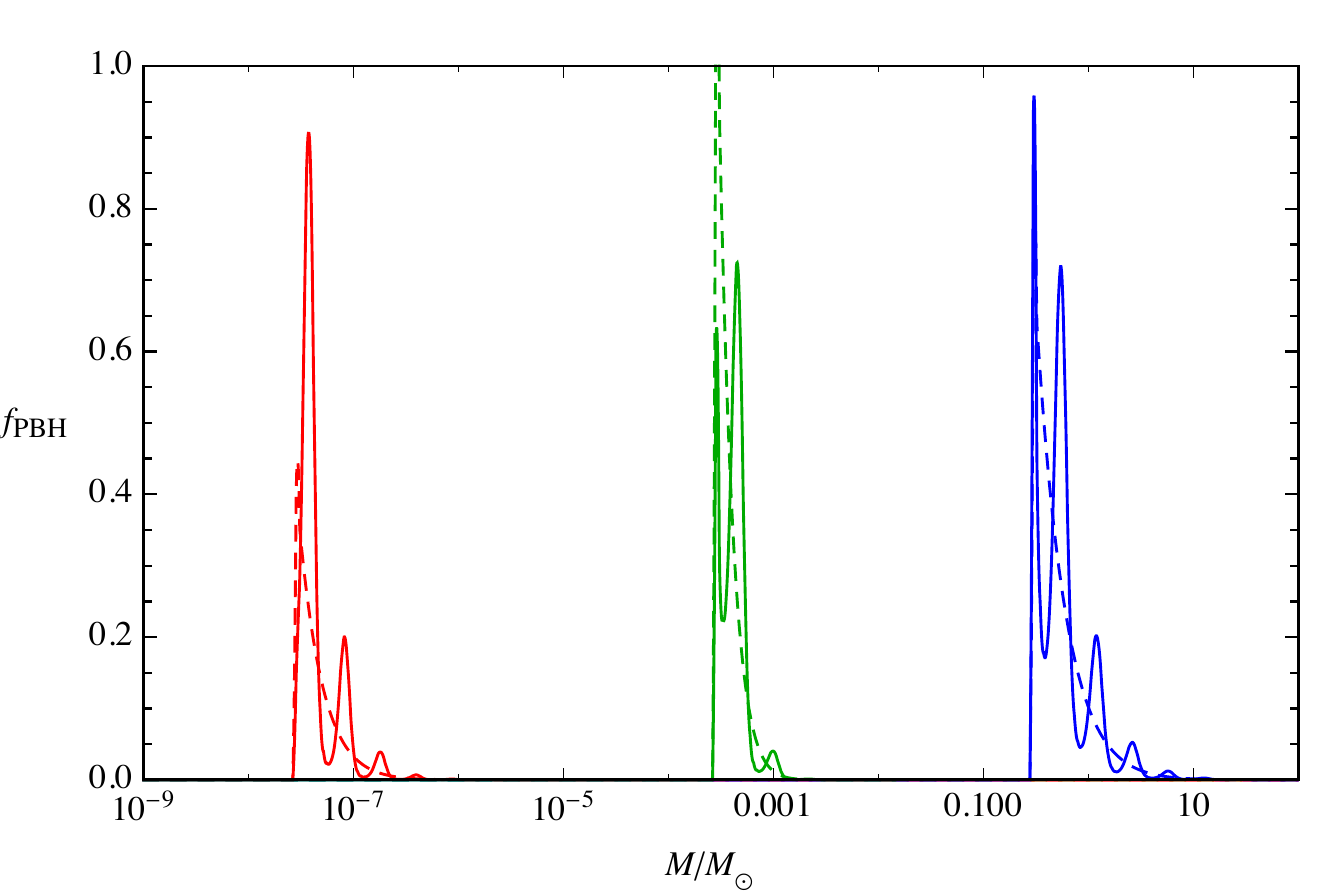}
	\caption{The present PBH dark-matter fraction $f_{\rm PBH}$ 
		for the axion-like curvaton inflation model as a function of $M / M_{\odot}$. 
		The parameters are (right to left) 
		${\widetilde\Pcal}^{\zeta}_{k_{\rm p}} = 1.06\times 10^{-2}$, 
		$M_{\rm min} = 10^{-3}M_{\frm}$, $\lambda = 2.1$ (red), 
		${\widetilde\Pcal}^{\zeta}_{k_{\rm p}} = 2.92\times 10^{-3}$, 
		$M_{\rm min} = 10^{-8}M_{\frm}$, $\lambda = 2.1$ (blue) and
		${\widetilde\Pcal}^{\zeta}_{k_{\rm p}} = 5.51\times 10^{-3}$, 
		$M_{\rm min} = 10^{-3}M_{\frm}$, $\lambda = 1.2$ (green). 
		The dashed curves show the unmodulated result, 
		while the solid ones include a particular choice of non-Bunch-Davies vacua.}
	\label{fig:Axion-like}
\end{figure}

The present dark-matter fraction as a function of mass is shown in Fig.~\ref{fig:Axion-like}, with the parameter choices indicated in the caption. The amplitude ${\widetilde\Pcal}^{\zeta}_{k}$ is chosen to produce a significant PBH density today. As before, the non-Bunch-Davies vacuum choice, together with the exponential enhancement of Eq.~\eqref{eq:fPBH-sim-int-erfc}, yields pronounced oscillatory spikes in $f_{\rm PBH}$, but only the first one is important since the model already contains spikes. When a minimum of the oscillation coincides with the spike of the standard mass function, the choice of non-standard vacua may lead to lowering of the peak in $f_{\rm PBH}$, as observed in the two middle (green) graphs in Fig.~\ref{fig:Axion-like}.

%%%%%%%%%%%%%%%%%%%%%%%%%%%%%%%%%%%%%%%%%%%%%%%%%%%%%%%%%
\subsection{QCD Phase Transition}
\label{sec:QCD-Phase-Transition}

\begin{figure}
	\centering
		\begin{picture}(0, 0)(0, 0)
			\put(-137, -98){\includegraphics[scale=0.55, angle=0]{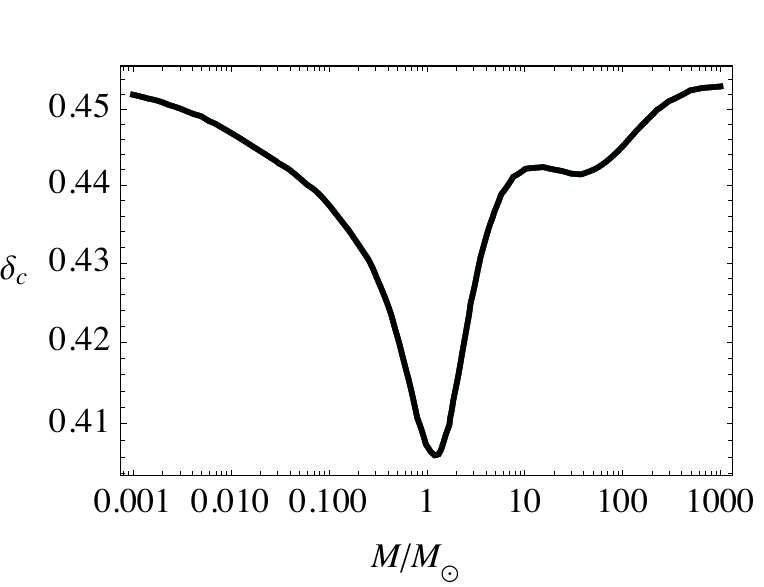}}
		\end{picture}\\[-3.2mm]
	\includegraphics[scale=1,angle=0]{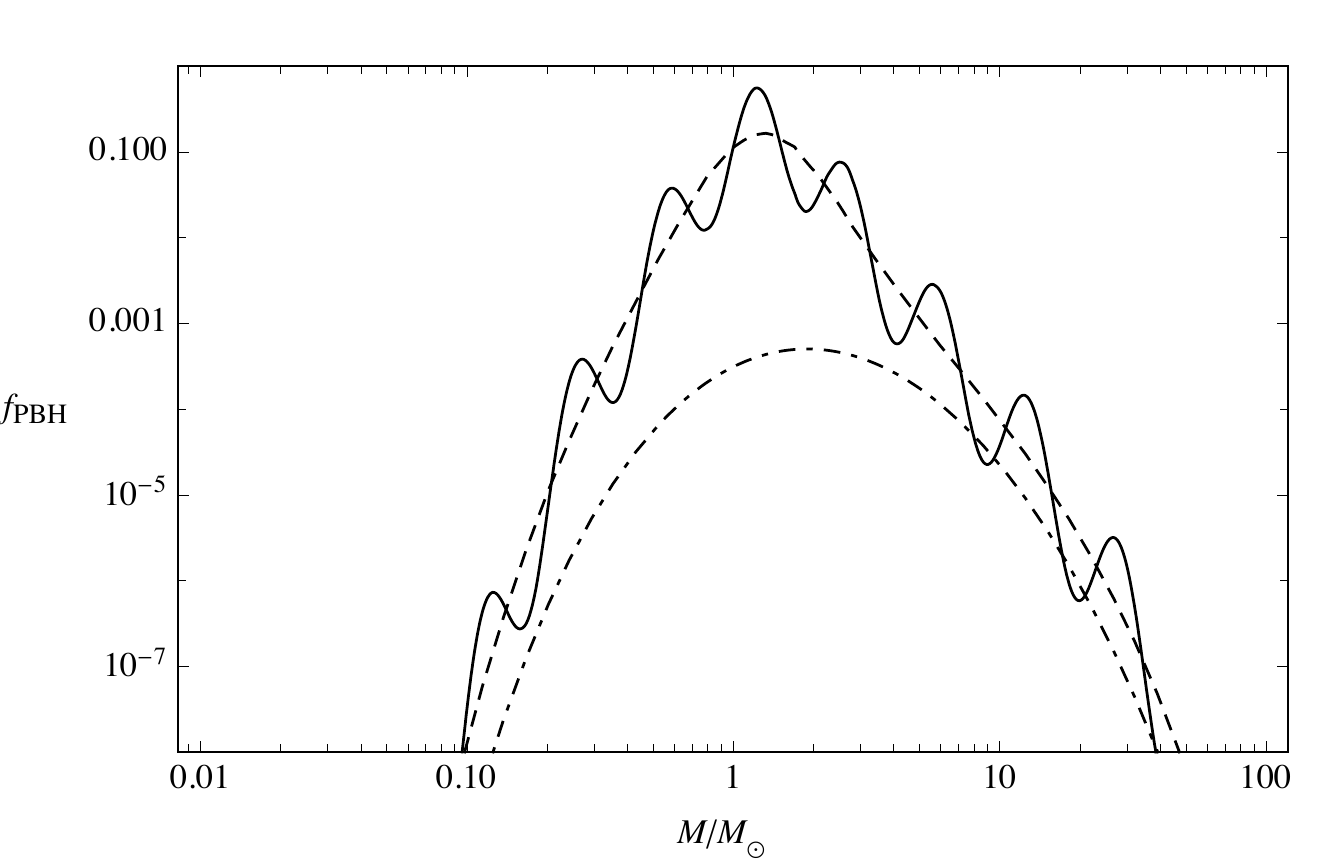}
	\caption{{\it Large figure:}~The present PBH dark-matter fraction $f_{\rm PBH}$ 
		during the QCD phase transition as a function of $M / M_{\odot}$. 
		The model assumed for the underlying power spectrum is 
		running mass inflation(dot-dashed curved, see Sec.~\ref{sec:Running-Mass}) 
		with $a = 0.011$, $b = 0.02$, $c = - 0.0023645$ and $\gamma = 0.36$.
		The dashed curves show the unmodulated result, 
		while the solid ones include a particular choice of non-Bunch-Davies vacua.
		{\it Small figure:}~form of the critical density perturbation 
		for collapse as a function of horizon mass 
		\cite{Byrnes:2018clq}.}
	\label{fig:QCD-Phase-Transition}
\end{figure}

\noindent Although the collapse process is more efficient during the QCD transition, due to the change in the equation of state (see Sec.~\ref{sec:QCD-Phase-Transition-II}), this enhancement is not strong enough to trigger PBH formation from inflationary perturbations with a non-running red spectrum. Hence some other enhancement of the power spectrum is required. For illustrative purposes, we use the running-mass model described in Sec.~\ref{sec:Running-Mass}, which has an increase in the power spectrum during the QCD phase transition. The present PBH dark-matter fraction as a function of mass is shown in Fig.~\ref{fig:QCD-Phase-Transition}, with the parameter choices being indicated in the caption. Pronounced oscillatory spikes can again be observed. The narrow feature associated with the QCD phase transition leads to an amplification at a mass scale of $\Ocal( 1 )\,M_{\odot}$.

%%%%%%%%%%%%%%%%%%%%%%%%%%%%%%%%%%%%%%%%%%%%%%%%%%%%%%%%%
\subsection{Scale Invariant Fluctuations}
\label{sec:Scale-Invariant}

\begin{figure}
	\centering
	\includegraphics[scale=1,angle=0, trim = 200 300 200 300]{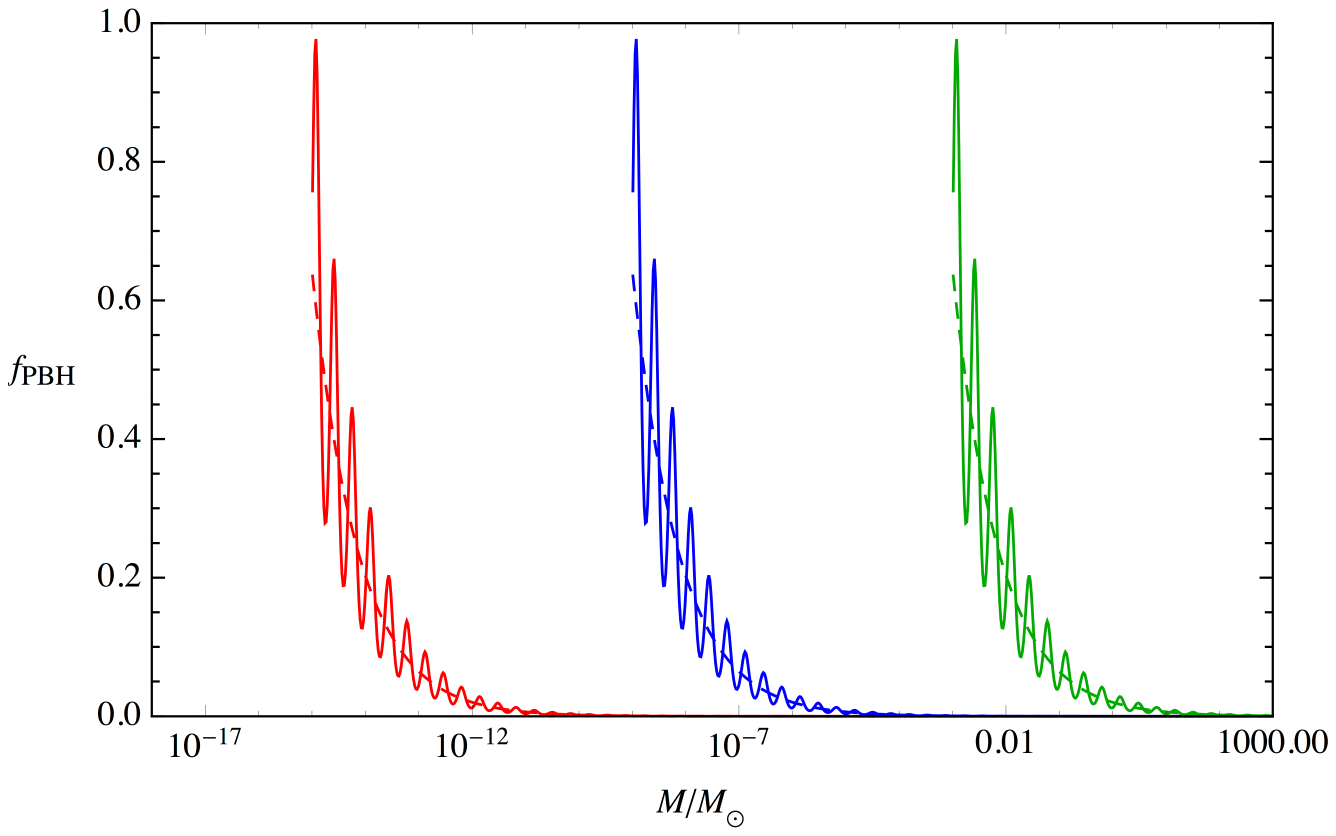}
	\caption{The present PBH dark-matter fraction $f_{\rm PBH}$ 
		for a scale-invariant power spectrum as a function of $M / M_{\odot}$. 
		In order to avoid overproduction of black holes at early times,
		a cut-off mass $M_{\rm cut}$ has been chosen at 
		$M_{\rm cut} = 10^{-15}\.M_{\odot}$ (red),
		$M_{\rm cut} = 10^{-9}\.M_{\odot}$ (blue), and
		$M_{\rm cut} = 10^{-3}\.M_{\odot}$ (green).
		The dashed curves show the unmodulated result, 
		while the solid lines include a particular choice of non-Bunch-Davies vacua.}
	\label{fig:Scale-invariant}
\end{figure}

\noindent For a scale-invariant spectrum, the mass PBH function in radiation-dominated era is proportional to $m^{-1/2}$ (see Sec.~\ref{sec:Collapse-from-Scale--Invariant-Fluctuations-or-Cosmic-Strings-II}). Hence, when superimposing the oscillatory features due to the non-Bunch-Davies choice of vacua, only one spike is likely to be important. This behaviour can be seen in Fig.~\ref{fig:Scale-invariant}, which shows $f_{\rm PBH}$ as a function of mass for the different choices of the cut-off mass $M_{\rm cut}$ indicated in the caption. As indicated by the linear scale of the y-axis in Fig.~\ref{fig:Scale-invariant}, the fall-off towards larger masses in this scenario is much milder than with the QCD phase transition. Therefore several spikes are likely to play a physical r{\^o}le if one of them does.\\[-5mm]

%%%%%%%%%%%%%%%%%%%%%%%%%%%%%%%%%%%%%%%%%%%%%%%%%%%%%%%%%
\section{Conclusions}
\label{sec:Conclusion-and-Outlook}

\indent In this paper we have discussed the effect of non-Bunch-Davies vacua on PBH formation and shown that these generally lead to PBH mass functions with multiple spikes. We have investigated this effect for a variety of scenarios: running-mass inflation, hybrid inflation, axion-like curvaton inflation, the QCD phase transition and scenarios with scale-invariant fluctuations. To the best of our knowledge, this constitutes the first derivation of such multimodal mass spectra from first principles. 

These scenarios could have interesting cosmological consequences. For example, multimodal mass spectra could contribute to the dark matter in mass ranges which would be precluded for single-peaked spectra, thereby allowing a higher PBH fraction of the dark matter. Also, having most of the PBH density in the relatively unconstrained range $10^{-13}$ -- $10^{-8}\,M_{\odot}$ still allows smaller peaks at larger masses. So a small peak around $10\,M_{\odot}$ could account for the LIGO black holes and an even smaller one at $10^{4}$ -- $10^{6}\,M_{\odot}$ could provide seeds for supermassive black holes in galactic nuclei.

As the PBH mass function depends exponentially on the primordial power spectrum, the oscillations due to the choice of non-Bunch-Davies vacua could easily lead to an overproduction of PBHs. This has important consequences for the allowed choice of model parameters. Also, if a minimum of the oscillatory onset in the power spectrum coincides with a spike in the standard mass function, the non-standard vacua could lead to a significant lowering of the peak in the PBH dark-matter fraction.

Although the amplitude, frequency and phase of the oscillations depend on the choice of vacua, there is certainly considerable freedom in assigning the PBHs different r{\^o}les. However, it should be stressed that the peaks with the non-Bunch-Davies vacua are equally spaced in logarithmic mass, so it generally hard to construct a scenario with more than two peaks at specific locations. For example, one could have peaks accounting for both the LIGO events and supermassive galactic black holes but not the dark matter as well. Or one could account for the dark matter and the LIGO events but not the supermassive black holes.\\

%%%%%%%%%%%%%%%%%%%%%%%%%%%%%%%%%%%%%%%%%%%%%%%%%%%%%%%%%
\acknowledgments

\noindent F.K.~acknowledges support by the Swedish Research Council (Vetenskapsr\r{a}det) through contract No.~638-2013-8993 and the Oskar Klein Centre for Cosmoparticle Physics. He also thanks the Queen Mary University of London for hospitality during parts of this project.

%%%%%%%%%%%%%%%%%%%%%%%%%%%%%%%%%%%%%%%%%%%%%%%%%%%%%%%%%
\setlength{\bibsep}{5pt}
\setstretch{1}
\bibliographystyle{utphys}
\bibliography{refs}

\end{document}